\documentclass[12pt,a4paper]{article}
 \usepackage{graphicx}
 \usepackage{afterpage}
 \usepackage{enumerate}
  \usepackage{longtable}
\usepackage{latexsym}
\usepackage{amsfonts}

\newcount\longrefs
\def\aap{\ifnum\longrefs=0 {\it Astron.\ Astrophys.}\else
{A\hbox{\rm \&}A}\fi}
\def\aapr{\ifnum\longrefs=0 {Astron.\ Astrophys.\ Rev.}\else
{A\hbox{\rm \&}AR}\fi}
\def\aaps{\ifnum\longrefs=0 {\it Astron.\ Astrophys.\ Suppl.}\else
{A\hbox{\rm \&}A Suppl.}\fi}
\def\aj{\ifnum\longrefs=0 {\it Astron.\ J.}\else
{AJ}\fi}
\def\ao{\ifnum\longrefs=0 {Applied Optics}\else
{Appl.\ Opt.}\fi}
\def\aspcs{\ifnum\longrefs=0 {\it Astron.\ Soc.\ Pacific Conf. Series}\else
{ASP Conf.\ Ser.}\fi}
\def\apj{\ifnum\longrefs=0 {\it Astrophys.\ J.}\else
{ApJ}\fi}
\def\apjl{\ifnum\longrefs=0 {Astrophys.\ J. Lett.}\else
{ApJ}\fi}
\def\aplett{\ifnum\longrefs=0 {Astrophys.\ J. Lett.}\else
{ApJ}\fi}
\def\apjs{\ifnum\longrefs=0 {\it Astrophys.\ J. Suppl.}\else
{ApJS}\fi}
\def\apss{\ifnum\longrefs=0 {Astrophys.\ and Space Science}\else
{Astrophys.\ Space Sci.}\fi}
\def\araa{\ifnum\longrefs=0 {Ann.\ Rev.\ Astron.\ Astrophys.}\else
{ARA\hbox{\rm \&}A}\fi}
\def\azh{\ifnum\longrefs=0 {\it Astronomicheskii Zhurnal}\else
{Astron.\ Zhur.}\fi}
\def\baas{\ifnum\longrefs=0 {Bull.\ Am.\ Astron.\ Soc.}\else
{BAAS}\fi}
\def\bain{\ifnum\longrefs=0 {Bull.\ Astronom.\ Institutes Netherlands}\else
{Bull.\ Astr.\ Inst.\ Neth.}\fi}
\def\gca{\ifnum\longrefs=0 {Geochim.\ Cosmochim.\ Acta}\else
{Geochim.\ Cosmochim.\ Acta}\fi}
\def\grl{\ifnum\longrefs=0 {Geophys.\ Res.\ Lett.}\else
{Geoph.\ Res.\ Lett.}\fi}
\def\iaucirc{\ifnum\longrefs=0 {IAU Circulars}\else
{IAU Circ.}\fi}
\def\ip{\ifnum\longrefs=0 {in press}\else
{in press}\fi}
\def\jgr{\ifnum\longrefs=0 {J.\ Geophys.\ Res.}\else
{J.\ Geophys.\ Res.}\fi}
\def\jrasc{\ifnum\longrefs=0 {J.\ Royal Astron.\ Soc.\ Canada}\else
{JRAS Can.}\fi}
\def\mnras{\ifnum\longrefs=0 {\textit {Mon.\ Not.\ Roy.\ Astron.\ Soc.}} \else
{MNRAS}\fi}
\def\nat{\ifnum\longrefs=0 {Nature}\else
{Nat}\fi}
\def\pasj{\ifnum\longrefs=0 {\it Pub.\ Astron.\ Soc.\ Japan}\else
{PASJ}\fi}
\def\pasp{\ifnum\longrefs=0 {\it Pub.\ Astron.\ Soc.\ Pacific}\else
{PASP}\fi}
\def\physscr{\ifnum\longrefs=0 {Physica Scripta}\else
{Phys.\ Scrip.}\fi}
\def\planss{\ifnum\longrefs=0 {Planetary \& Space Science}\else
{Plan. \& Space Sci.}\fi}
\def\procspie{\ifnum\longrefs=0 {Proc.\ SPIE}\else
{Proc.\ SPIE}\fi}
\def\qjras{\ifnum\longrefs=0 {Quarterly J.\ Royal Astron.\ Soc.}\else
{QJRAS}\fi}
\def\sa{\ifnum\longrefs=0 {\it Soviet Astron..}\else
{Sov.\ Astron.}\fi}
\def\skytel{\ifnum\longrefs=0 {Sky \& Telescope}\else
{Sky \& Tel.}\fi}
\def\solphys{\ifnum\longrefs=0 {\it Solar Phys.}\else
{Sol.\ Phys.}\fi}
\def\ssr{\ifnum\longrefs=0 {\it Space Science Rev.}\else
{Space\ Sci.\ Rev.}\fi}

\begin{document}
%%=========================\swarrow=================================
\small

 \title{\bf Convective Line Shifts in the Spectra of Solar-Type Stars
}
\author{\bf V.A. Sheminova }
  \date{}

 \maketitle
 \thanks{}

\normalsize

\begin{center}
{Main Astronomical Observatory, National Academy of Sciences of Ukraine,\\ Akademika  Zabolotnoho 27,  Kyiv,  03143, Ukraine,\\ e-mail: shem@mao.kiev.ua
}
\end{center}

\normalsize

\begin{abstract}
The Doppler line shifts in the spectra of the Sun and stars with effective temperatures from 4800 to 6200 K were measured and the average connective (granulation) velocities were estimated. The absolute scale of the line shifts for the stars was established on the basis of the derived dependence of the shifts of solar lines on optical depth. For FGK solar-type stars, curves of convection velocities as a function of the height in the atmosphere in a large range of heights from 150 to 700 km were obtained for the first time. All these curves indicate a decrease in blue shifts with height, which means that the granulation velocities through the photosphere slow down to zero. In the lower chromosphere, red shifts of strong Mg I lines are observed, which indicate a change in the direction of granulation velocities to the opposite and confirm the effects of reversal of granulation at heights above 600 km. In cooler K stars, granulation shifts change with height on average from $-50$ to 100 m/s, while they change more sharply in hotter FG stars from $-700$ to 300 m/s. The gradient of the line shift curves increases with an increase in the effective temperature and a decrease in gravity, metallicity, and age of the star. The connective velocity of the star averaged over all analyzed heights increases from $-90$ to $-560$ m/s from colder to hotter stars. It correlates with macroturbulence, asymmetry of spectral lines, and the rotation velocity of the star. We also obtained the radial velocities of the stars and compared them with the SIMBAD data. Large deviations of $-21 050$ and 1775 m/s were found for the stars HD 102361 and HD 42936, respectively. For the rest of the stars, the deviation does not exceed $\pm340$ m/s, which is probably associated with the use of an average granulation velocity of $-300$ m/s in the SIMBAD data. Our analysis has shown that the average granulation velocity is not the same for solar-type stars. It is lower in colder stars and higher in hotter stars than the Sun. Therefore, determination of the radial velocities needs to take into account the individual granulation velocities of stars.

\vspace {0.5cm}
{\bf Keywords:}  solar-type stars, granulation, line shifts, connective velocities, radial velocities

\end{abstract}

\section{Introduction
}

The granulation region in stellar atmospheres is the region where convective flows leave the connective envelope. The formation and evolution of stellar granulation are described in detail by Dravins \cite{1990IAUS..138..397D}. The ascending flows create a granulation pattern that is clearly visible only on the Sun's surface and not visible on the surface of stars due to low resolution. The emerging hot matter is called granules, and the sinking matter is called intergranules. Since the granules dominate, the observed spectral lines of the stars exhibit a Doppler blue shift. The velocity of convective flows decreases with height in the photosphere, so the cores of strong lines, which form in higher photospheric layers, undergo small blue shifts. The weaker the spectral line, the deeper the layers in which its core forms, and the greater the blue shift it undergoes. In general, connective motions penetrating the photosphere broaden the line profiles, make them asymmetric, and shift the central position of the lines in comparison with a static atmosphere. These changes in the lines were named the first, second, and third signs of granulation in stellar atmospheres \cite{2009ApJ...697.1032G}.

The first sign, or the broadening of the profiles, is mainly characterized by the velocities of motions caused by the rise and fall of matter in the granulation structures. The velocity of large-scale movements, called macroturbulent velocity, changes the shape of the profile along the line of sight without changing the equivalent line width. In FGK solar-type stars, the macroturbulent velocity decreases with height in the atmosphere, its gradient increases, and its average value grows from 1.7 to 6.0 km/s with an increase in effective temperature and luminosity  \cite{2010ApJ...710.1003G, 2019KPCB...35..129S}. It is believed that the macroturbulent velocity derived from the profiles of the observed lines is actually a rough measure of the variance of convective (or granulation) line-of-sight velocities in the photosphere relative to their mean value \cite{2007IAUS..239..103L}. The velocity of granulation movements on scales smaller than the average free path of a photon in a spectral line is called microturbulent velocity; it broadens the line profile and changes the equivalent width. This velocity is small in value and decreases slightly with height. In solar-type stars, it grows with effective temperature from 0.7 to 1.5 km/s \cite{2019KPCB...35..129S}. The line profile is also affected by the rotation velocity of the star $v\sin i$, which broadens the profiles without changing the equivalent width. As a rule, it can be detected if it exceeds 2--3 km/s. Otherwise, the broadening due to rotation can mask the macroturbulent effects. Other movements, such as nonradial oscillations, have little effect on the shape of the profiles.

The second sign of granulation, asymmetry of line profiles, is characterized by a bisector, the shape of which reflects variations of the shifts in the line profile with height in the photosphere, since the height of formation of each point of the profile increases from the distant wing of the profile to its core. The predominant motion of the granules, the velocity of which changes with height, determines the asymmetry of the line profiles in the granulation region. In solar-type stars, the shape of the bisector in strong lines looks approximately like the letter ``C'' with the maximum deviation toward the blue side, while the bisector resembles the upper part of the letter ``C'' in weak lines, and it looks like ``/'' in stars with rotation velocities $v \sin i > 5$ km/s \cite{2002ApJ...566L..93A, 2018ApJ...857..139G, 1999PASP..111.1132H, 2020KPCB...36..291S}. The span of the bisector increases with the effective temperature and indicates an increase in the force of convection \cite{2010ApJ...721..670G, 2020KPCB...36..291S}. Starting with main-sequence stars F0 V and higher, the shape of the bisector is reversed. The transition from one form to another was called the ``granulation boundary'' \cite{1986PASP...98..499G}. This phenomenon is explained by a different pattern of granulation in hot stars: slow, cold downflows of matter that occupy a large area on the surface versus hot, faster, and narrower upflows \cite{1990ASPC....9...27D, 2010ApJ...721..670G, 1998A&A...338.1041L}.

The third sign is a convective or granulation line shift, which indicates the average velocity of ascending and descending convective flows at the depth of effective formation of the line core. Since spectral lines form at different heights in the atmosphere, it is possible to trace the variation in the magnitude of the shifts with height. Granulation shifts are well studied for the Sun and some stars \cite{2002ApJ...566L..93A, 1999ASPC..185..268D, 2009ApJ...697.1032G, 1999PASP..111.1132H, 2008A&A...492..841R, 2009A&A...501.1087R}. The use of the third sign has one advantage, which is the possibility to involve a significantly larger number of lines for analysis than in the case of the first and second granulation signs. Blend-free line cores are more common in spectra than blend-free profiles. A large number of lines will significantly reduce the error of the result. On the other hand, it is rather difficult to investigate line shifts since we encounter an uncertainty in measuring the center wavelength of the line, insufficient spectral resolution and low signal-to-noise ratio in stellar spectra, and insufficiently high accuracy of laboratory wavelengths, gravitational red shift, and radial velocity of the star relative to the observer. In addition, low spatial resolution of stellar spectra does not allow one to take into account variations from the center of the disk to the limb and to trace the time variations in the shifts that occur during the evolution of granulation and cyclic magnetic activity, which weakens local convective motions.

Additionally, a fourth sign of granulation, flow deficit, was introduced \cite{2018ApJ...857..139G}. This parameter is a measure of the flow contrast between granules and intergranules. Based on the analysis of this parameter, it was concluded that the well-known granulation boundary in stars \cite{1986PASP...98..499G} does not indicate in which stars convection affects or does not affect the field of photospheric velocities; it instead shows the relationship between the magnitude of the velocities and their gradients across the photosphere \cite{2010ApJ...710.1003G, 2010ApJ...721..670G}.

At the present time, we still know little about the granulation velocities of stars. Their accurate values are necessary to solve the existing problem of measuring radial velocities, for example, in problems of searching for exoplanets \cite{2008A&A...492..199D, 2002A&A...390..383G, 2002ApJS..141..503N, 2008A&A...492..841R}. According to the conclusions of \cite{2017chsw.confE..15C}, granulation should be considered as an internal uncertainty in accurate measurements of exoplanet transits; therefore, the complete characterization of granulation is important to determine the degree of uncertainty in the planet's parameters. In this regard, we set ourselves the task of studying the granulation of solar-type stars in more detail: measure connective shifts of spectral lines, determine average granulation velocities in a number of stars with different parameters, investigate variations in granulation velocities with height in the atmosphere, and trace possible tendencies of connective shifts depending on the main parameters of stars.

This study is organized as follows. Section 2 presents a brief overview of the main results obtained up to this time. Section 3 describes the observational data, parameters of the analyzed stars, measurements of the central position of the line in the stellar spectra, and calculation of the convective line shifts. Our results, as well as the discussion of our findings, their scientific significance and comparison with literary estimates, are presented in Section 4. Section 5 contains our main conclusions.

\section{Brief overview of the results}

{\bf Solar shifts.} A systematic analysis of line shifts in the integral spectrum of the Sun is of particular importance as a reference for the study of stellar granulation. A wide range of shifts of various Fraunhofer lines has been obtained by this time \cite{1998A&AS..129...41A, 2000A&A...359..729A, 1984SoPh...93..219B, 1999ASPC..185..268D, 1981A&A....96..345D, 1999PASP..111.1132H, 2005SoPh..196...41P}. The magnitude of these shifts depends on the strength of the lines, excitation potential, and wavelength. All authors note that the dependence of the shifts on the line depth is the most significant. The blue shift values approach a constant close to zero with increasing line strength and form a ``plateau'' for strong lines with an equivalent width of more than 200 mA, which occurs due to the saturation of strong lines \cite{1998A&AS..129...41A}. The dependence of the shifts on the excitation potential turned out to be apparent since it reflects the sensitivity to the line depth, and the dependence on the wavelength exists only for lines with the same depths \cite{1999PASP..111.1132H}. The shorter the wavelength, the greater the blue shift of the line due to the attenuation of continuous absorption with decreasing wavelength. On the basis of the new IAG atlas with a resolution of $R = 1 000000$, an expression was obtained in \cite{2016A&A...587A..65R} for the magnitude of the shift $V$ (in m/s) depending on the depth $d$ of the line:
\[ V = -504.891-43.7963d -145.560d^2 + 884.308d^3. \]
According to this expression, the difference between the granulation shifts of weak and strong lines is $-550$ m/s on average. Slightly later, Gray \cite{2018ApJ...857..139G} obtained a new expression based on his own observations ($R = 100 000$):
\[ F/Fc=0.3249 - 1.2919\cdot 10^{-4}V + 5.8390\cdot 10^{-7}V^2 -  2.0638\cdot 10^{-9}V^3,\]
as well as using the atlas \cite{2011ApJS..195....6W} ($R=700000$):
\[ F/Fc=0.3056 - 2.9902\cdot 10^{-4}V + 3.2008\cdot 10^{-7}V^2 -  2.1689\cdot 10^{-9}V^3,\]
where  $F/F_c$ is the ratio of the line flux to the continuum flux. These expressions were used to scale the observed stellar shifts \cite{2012AJ....143...92G}.

The magnitude of the line shifts is significantly affected by solar activity. Blue shifts are weakened in magnetic regions due to changes in the granulation structure \cite{1990A&A...231..221B, 1997KPCB...13e..65B}. The shifts are also affected by the spectral resolution, which leads to a change in the slope of the dependence of the shifts on the line depth \cite{1999PASP..111.1132H}.

An important aspect in measuring granulation shifts is the correctness of the absolute scale of shifts. The absolute scale is understood as a scale of shifts that does not contain the gravitational shift and is corrected for the Doppler shift due to the relative motion of the Earth and the star along the line of sight. The scales of different solar atlases were compared in \cite{2016A&A...587A..65R}. It turned out that the scale of Kurucz's atlas \cite{1984sfat.book.....K} is shifted by $-50$ m/s in the range of 400--500 nm, while the scale for the IAG atlas \cite{2016A&A...587A..65R} is consistent over the entire wavelength range and is accurate within the uncertainty of the position of telluric lines, i.e., with an accuracy of 10 m/s. The scale for the solar spectrum in the atlas by Hinkle et al. \cite{2005ASPC..336..321H} is shifted by $-100$ to $-330$ m/s in the range of 450--650 nm, as shown in \cite{2020KPCB...36..291S}. The HARPS LFC atlas \cite{2013A&A...560A..61M} with $R\approx120000$ is currently the most accurately calibrated in terms of wavelength, apart from the IAG atlas.

{\bf Stellar shifts.} The measurement of granulation line shifts in stellar spectra is much more complicated due to insufficient spectral resolution, increased noise, and the existing problem of the absolute scale of shifts. The obtained results show a very large spread in the shifts not only for weak lines but also for all other lines
\cite{2002ApJ...566L..93A, 1999ASPC..185..268D, 2009ApJ...697.1032G, 1999PASP..111.1132H, 2007IAUS..239..103L, 2017A&A...607A.124M, 1989ASIC..263..125N, 1988ApJ...327..321N,  2011A&A...526A.127P, 2008A&A...492..841R, 2009A&A...501.1087R}.
%(Nadeay, Maillard (1988), Nadeau, Bedard, Maillard (1989),  Дравинс (1999), Альенде Прието  (2002), Hamilton \& Lester 1999; Landstreet 2007;  Ramirez et al. (2008), Gray 2009,  Pasquini et al. (2011), Ramirez et al. 2009, 2010), Meunier et al. 2017)).
While studying this problem, Gray  \cite{2009ApJ...697.1032G} noted a number of causes of the large spread for weak lines: (1) large uncertainty in the measurement of the central wavelength of the line due to the flatter bottom of the profile as compared to moderate lines, (2) distortion by weak unidentified blends, (3) large errors of laboratory wavelengths in the data
\cite{1994ApJS...94..221N},% Nave et al. (1994)
and (4) the use of lines in a wide range of wavelengths. In addition, the accuracy of shift measurements in stars is additionally influenced by the rotation velocity of the star. To reduce the large spread, the shift is measured as an average over a large number of lines with similar conditions of their formation, i.e., with close effective heights of line formation.

As noted by many researchers, the main property of granulation line shifts in stars is their growth with an increase in effective temperature due to stronger surface convection. The granulation effects in K dwarfs are smaller than those of the Sun
\cite{1987A&A...172..211D, 1982ApJ...255..200G}. % (Dravins 1987a; Gray 1982).
 The convective line shifts measured in the spectra of K giants from weak and strong lines differ by almost 1 km/s
\cite{2002ApJ...566L..93A}. %(Альенде Прието  (2002, 566) ).
This is significantly more than for GK dwarfs and subgiants of similar spectral types. For red giants, the shifts vary in the range from $-100$ to $-900$~m/s
\cite{2010ApJ...725L.223R}. %(Ramirez et al. (2010)).
For stars from K2 to G0, blue shifts increase from $-150$ to $-500$~m/s and decrease with decreasing stellar mass and increasing metallicity
\cite{2017A&A...597A..52M, 2017A&A...607A.124M}. % (Meunier et al. 597, (2017), Meunier et al. 607, (2017)).
The slope of the dependences of the shifts on the line strength varies in different stars
\cite{2009ApJ...697.1032G}.  %(Gray (2009).
The magnitude of stellar shifts depends on magnetic activity 
\cite{1992ApJ...400..681G, 1996ApJ...465..945G, 1988ApJ...327..399W}.
%(Wallace, Huang \& Livingston (1988), Gray et al., 1992,400; 1996, 456; 465).
Since the area occupied by magnetic fields on the stellar disk varies over the cycle of stellar activity, the line shifts are modulated with the same period. The results of
 \cite{2017A&A...607A.124M}  % Meunier et al.607, (2017)
also showed the attenuation of convection with stellar activity in the range of effective temperatures 6300--5200 K for stars from F7 to K4.

%%%%%%%%%%%%%%%%%%%%%%%%%%%%%%%%
%%==========================================================

%___________________________________ Table 1
\begin{table}
\centering
 \caption{\small Parameters of the analyzed stars: effective temperature  $T_{\rm eff}$, surface gravity  $\log g$, metallicity [M/H], mass
 $M/M_\odot$, age $t$, radius $r$, luminosity  $L$, and iron abundance  $A$}.
 \vspace {0.3 cm}
\label{T:2}

\footnotesize
\begin{tabular}{lccccccccc}
\hline\hline
HD&Type & $T_{\rm eff}$, K&$\log g$&[M/H]&$M/M_\odot$   &$r$ & $L$  &$t$  &$A$     \\
  &     &             &        &     &              &    &      &($10^9$ years)&        \\
\hline
6790  & G0 V      &6012   &4.40   &$-$0.06  & 1.089& 1.13& 1.427&  3.5 & 7.55   \\
38459 & K1 IV-V   &5233   &4.43   &~~0.06   & 0.882& 0.84& 0.515&  9.0 & 7.58   \\
42936 & K0 IV-V   &5126   &4.44   &~~0.19   & 0.881& 0.91& 0.510& 12.0 & 7.61   \\
93849 & G0/1 V    &6153   &4.21   &~~0.08   & 1.268& 1.52& 2.974&  3.5 & 7.66   \\
102196& G2 V      &6012   &3.90   &$-$0.05  & 1.395& 1.97& 4.719&  3.0 & 7.52   \\
102361& F8 V      &5978   &4.12   &$-$0.15  & 1.250& 1.54& 2.775&  2.0 & 7.39   \\
127423& G0 V      &6020   &4.26   &$-$0.09  & 1.107& 1.16& 1.546&  3.1 & 7.48   \\
128356& K2.5 IV   &4875   &4.58   &~~0.34   & 0.824& 0.84& 0.371& 15.5 & 7.73   \\
147873& G1 V      &5972   &3.90   &$-$0.09  & 1.493& 2.36& 6.568&  2.6 & 7.53   \\
158469& F8/G2 V   &6105   &4.19   &$-$0.14  & 1.223& 1.42& 2.498&  2.0 & 7.41   \\
189627& F7 V      &6210   &4.40   &~~0.07   & 1.244& 1.43& 2.719&  4.0 & 7.67   \\
221575& K2 V      &5037   &4.49   &$-$0.11  & 0.823& 0.82& 0.368&  6.0 & 7.34   \\
Sun~~~& G2 V      &5777   &4.44   &~~0.00   & 1.000& 1.00& 1.000&  4.6 & 7.52   \\
\hline
\end{tabular}
\end{table}
\noindent

\section{Initial data and method of measurements}

{\bf Observations.} We use the spectra of 12 solar-type stars (Table 1) from a sample of Calan-Hertfordshire Extrasolar Planet Search (CHEPS), which were obtained at HARPS in La Silla in Chile
\cite{2009MNRAS.398..911J} %(Дженкинс и др. 2008; 2009; 2011)
with a resolution  $R \leq 120000$ and a signal-to-noise ratio  $>100$. The data on effective temperatures $T_{\rm eff}$, parameters of gravity  $\log g$  and metallicity [Fe/H] were taken from
\cite{2017MNRAS.468.4151I}, %Иванюка и др.  (2018)
the data on the mass $m$ and age from
             \cite{2019A&A...621A.112P}, %Павленко и др.(2019). )
  the data on micro- and macroturbulent velocities, rotation parameters $v \sin i$ and iron abundance $A$ from 
 \cite{2019KPCB...35..129S}. % Sheminova.
The radii and line-of-sight velocities of the stars correspond to the data from the Gaia DR2 catalog
\cite{2018A&A...616A...1G}. %(Gaia Collaboration, 2018, )
The spectral types correspond to data from the SIMBAD database (http://cdsportal.u-strasbg.fr). All the stars are single and inactive ($\log R_{\rm HK} \leq -4.5$ dex).
The $T_{\rm eff}$  range is small, from 4800 to 6200 K. Some of the stars are slightly rich in metals with [Fe/H] from 0.06 to 0.34, while others have a minor metal deficit from $-0.05$ to $-0.15$. We also used the solar integral flux spectrum obtained at HARPS \cite{2013A&A...560A..61M}. This atlas was calibrated using an ideal laser frequency comb (LFC) calibrator; therefore, the wavelength solution is the most accurate currently available. In addition, we used the IAG solar flux atlas
\cite{2016A&A...587A..65R} %(Reiners et al. (2016))
with a resolution  $R\approx 1000 000$, the zero point of which was set with an accuracy of 5--10 m/s.

{\bf Measuring the central wavelength of a spectral line.} There are various methods for measuring the central wavelength of a line, and the details of these methods affect the accuracy of the results. The most popular methods include Gaussian and polynomial approximation
\cite{1998A&AS..129...41A, 2002ApJ...566L..93A,1984SoPh...93..219B, 2008A&A...492..841R, 2016A&A...587A..65R}.
%Balthasar (1984),  Allende Prieto \& Garcia Lopez 1998, 2002, 566, Ramirez (2008), Reiners (2016).
The use of polynomials turned out to be quite effective at high resolution. Aside from that, a bisector point 7\% above the minimum of the line
\cite{1987A&A...172..211D}, %(Dravins 1987,172),
the center of gravity of the line core 
\cite{1997KPCB...13e..65B}, % (Гадун и Шеминова),
the two lowest points of the bisector in the line core
\cite{1999PASP..111.1132H}, % (Hamilton, Lester (1999)),
and the lowest point of the bisector that corresponds to one data point to the left of the minimum
\cite{2018ApJ...857..139G} % (Gray 2018,853).
are also used. In most methods, the measurement accuracy depends on the strength of the line and the width at the line core.

To select the most noise-resistant method for our particular observed spectra, we tested most of the known methods. The observed line profiles were smoothed to reduce noise and interpolated. The width of the region around the core of the line was selected depending on the star and was 3--7 km/s. When the center of gravity was used, the optimal region corresponded to the part of the core at a height of 5--7\% above the minimum. To represent the test results in the convective velocity scale, we measured the shift of the observed line relative to the laboratory wavelength and subtracted the shift calculated based on the radial velocities from the SIMBAD base and the gravitational shift calculated using the mass
\cite{2019A&A...621A.112P}  %(Pavlenko
and radius
\cite{2018A&A...616A...1G} of the star.
%(Gaia DR2 (Gaia Collaboration, 2018)).

Figure 1 shows polynomial approximations of the granulation line shifts obtained for each star using different methods of measuring the line wavelength. All these shift curves are shown on the same scale. The shifts for two stars, HD 102361 and HD 42936, were found to be unrealistic. In all likelihood, this was caused by the errors in the radial velocities used from the SIMBAD database. As seen from Fig. 1, weak lines ($d<0.2$) show the greatest differences between the curves obtained by different methods, due to the higher sensitivity to blends and noise. The same is observed for strong lines ($d > 0.8$) in stars with high rotation velocities, mainly due to the expanded and flat core of the line and the small number of strong lines in these stars. The two-point method turned out to be the most sensitive to noise and to a flat core of the line. For the hottest star HD 189627, the differences in the shifts of strong lines are especially large. To estimate the reliability of the method, we compared the standard deviations of the shifts for each star averaged over all bins. It turned out that the best result is given by the parabola   ($<\sigma>=100$--180 m/s), while the worst result is given by the Gaussian and the two-point method  ($< \sigma>= 150$--260 and 120--300 m/s respectively). The width of these ranges shows the variation in the $< \sigma>$ values from cold to hot stars. The largest $< \sigma>$ value was obtained for weak lines in hot stars HD 189627, HD 147873, and HD 102361 with a rotation velocity $v \sin i>5$  km/s. Based on these tests, it was concluded that a polynomial of the second degree is the most reliable for measuring the wavelengths of lines in the spectra of stars in our sample.

%%%%%%%%%%%%%%%%%%%%%%%%%%%%%%%%%%%%%%%%%%% Figure 1
 \begin{figure}[!t]
\centerline{
\includegraphics [scale=1.1]{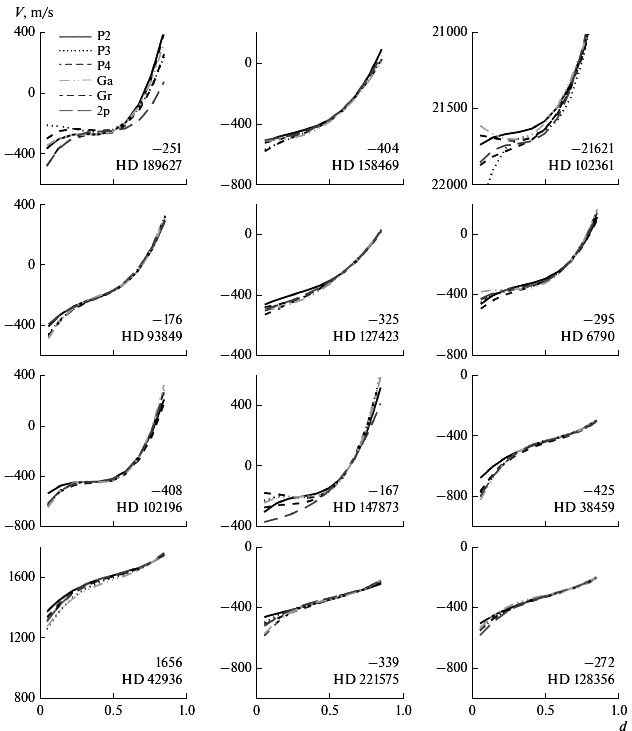}
     }
 \caption {\small
 Testing of the methods for measuring the central position of a line in the spectrum of solar-type stars. Each curve is a polynomial approximation of convective shifts $V$ versus line depth $d$ obtained by different methods using polynomials of the second degree (P2), third degree (P3), fourth degree (P4), Gaussian (Ga), center of gravity (Gr), and two points closest to the center of the line (2p). Each panel shows the average granulation shift in m/s.}
\label{prof1}
 \end{figure}
%%%%%%%%%%%%%%%%%%%%%%%%%%%%%%%%%%%%%%%%%%%

{\bf Measurement of the granulation velocity.}
We measured the central wavelength  $ \lambda_{\rm obs}$  of the line in the stellar spectrum using a parabola and obtained the observed shift using the well-known formula for the Doppler shift in m/s: 
 $V_{\rm obs} = c(\lambda_{\rm obs} - \lambda_{\rm lab}){/}\lambda_{\rm lab}$, where $\lambda_{\rm lab}$ is the laboratory wavelength, and $c$ is the speed of light. To obtain the granulation velocity at the depth of formation of the core of a given line, it is necessary to subtract the gravitational shlft $V_{\rm grav}$  and the velocity of the relative motion of the Earth and the star along the line of sight, i.e., the radial velocity  $RV$ from $V_{\rm obs}$. Velocity $V_{\rm grav}$ was calculated using the formula $V_{\rm grav}= (M_{\rm {star}} {/}M_{\odot})(R_{\odot}  {/}R_{\rm {star}}) V_{\rm grav, \odot}$, where $M_{\rm {star}}$ and $R_{\rm {star}} $ are the mass and radius of the star, and $V_{\rm grav, \odot}= 633$  m/s is the solar gravitational shift according to
\cite{1999ASPC..185...73L}. %(Lindegren et al. 1999),
The test results (Fig. 1) with the use of radial
velocities from the SIMBAD database showed that the obtained average convective shifts in the stars do not agree with the main parameters of the stars; in particular, they do not show a dependence on the effective temperature. This can be seen in Fig. 1, where the temperature of the stars increases from left to right and from top to bottom. The modern data on the radial velocities are obtained by spectroscopic and astrometric methods. The latter make it possible to obtain radial velocities that do not depend on such phenomena as pulsations, convection, rotation, wind, isotopic composition, pressure, and gravitational
potential
\cite{2005ESASP.560..113D}. % (Dravins, L. Lindegren (2005).
Unfortunately, they are not known for all stars. Spectroscopic measurements give less accurate radial velocities due to poorly known convective velocities. Since the problem of radial velocities has not been solved completely, one often measures either differential convective line shifts, for example, shifts of weak lines relative to strong lines in the spectrum of the same star or shifts relative to a selected reference point. In magnetic formations of the solar photosphere, the very strong line Mg I 517.27 nm served as a reference for measuring convective shifts
\cite{1997KPCB...13e..65B, 1983IAUS..102..149L}. % Livingston (1983), Гадун и Шеминова (1997).
 It was assumed that its core forms high in the atmosphere, where the correlation between the intensity and velocity of convective motions is weak, so its convective shift will be almost zero. The group of strong iron lines is also used as a reference 
\cite{1998A&AS..129...41A, 1997KPCB...13e..65B, 2008A&A...492..841R}
%(Гадун и Шеминова (1997), Альенде Прието и Гарсия Лопес (1998a, b), Ramires 2008).
or the absolute scale of line shifts in the solar flux spectrum obtained with high-precision wavelength calibration
\cite{2009ApJ...697.1032G, 2017A&A...597A..52M}. %Gray (2009), Meunier,A\&A 597, A52 (2017).

The difficulty of this task also lies in the fact that we do not have high-resolution spectra with accurate wavelength calibration and precise laboratory wavelengths for stars. The most accurate laboratory line wavelengths to date are presented in the NIST database (https://physics.nist.gov/asd), which contains corrected values for iron lines
\cite{1994ApJS...94..221N}. % Nave et al. (1994)
These data were used in our analysis.

To determine the zero-point of the absolute scale of line shifts in the spectra of stars with unknown radial velocities, we took the absolute shifts of solar lines as a standard. The calculation scheme was as follows.

(1) Calculate the effective formation depths of line cores for the Sun and all stars.

(2) Measure the granulation line shifts for the Sun in the absolute scale, which takes into account the gravitational shift and the radial motion of the star, and construct the graph of the resulting shifts depending on the effective depths of formation of the lines.

(3) Determine the optical (or geometrical) depth at which the granulation shifts of the solar lines are zero and assume that it is the same for all solar-type stars. This depth is the zero point of the absolute scale of shifts for each star. We will call it the depth of zero shifts.

(4) Measure the shifts of the observed lines$V_{\rm obs}$ in stars and plot their graphs depending on the effective formation depth of these lines for each star.

(5) Move the scale of stellar shifts so that the zero shifts correspond to the depth of the zero shifts. The use of this scheme allows us to set the zero point of absolute shifts and obtain the granulation shift $V$ for each star as well as calculate the radial velocity $RV$ of the star by subtracting the gravitational shift of the star $V_{\rm grav}$ from the displacement of the stellar shift scale.

The effective optical depths of formation of line cores $\log \tau_5$ were calculated based on the depression contribution functions  \cite{2015arXiv150500975G, 1974SoPh...37...43G}. For this purpose, we have performed the synthesis of absorption lines in stellar spectra in the LTE approximation using the SPANSAT code \cite{1988ITF...87P....3G} and MARCS model atmospheres \cite{2008A&A...486..951G} with the values of the main parameters of the stars, the parameters of the micro- and macroturbulent velocities, the rotation velocity $v\sin i$, and the iron abundance $A$ (presented in Table 1). The oscillator strengths of the iron lines were taken from the tables
\cite{2006JPCRD..35.1669F}. %Fuhr, Wiese (2006).
The damping constant associated with the van der Waals force between the absorbing and perturbing atoms is  $\gamma_6$  in accordance with the classical Unsold approximation. Our assumptions are acceptable for estimating the average depth of formation of spectral absorption lines in stellar atmospheres. The list of analyzed iron lines was selected to cover a wide range of line depths for each specific star. In addition, three very strong Mg I lines (517.26, 518.36, and 552.84 nm) and two Ca I lines (612.2 and 616.2 nm) were used to broaden the red shift range. Based on the idea that granulation affects lines that form in a narrow atmospheric layer in the same way, all lines were divided into groups with similar formation depths. Thus, a statistically reliable result was obtained.

%%%%%%%%%%%%%%%%%%%%%%%%%%%%%%%%%%%%%%%%%%% Figure 2
 \begin{figure}[t]
 \centerline{
 \includegraphics   [scale=1.1]{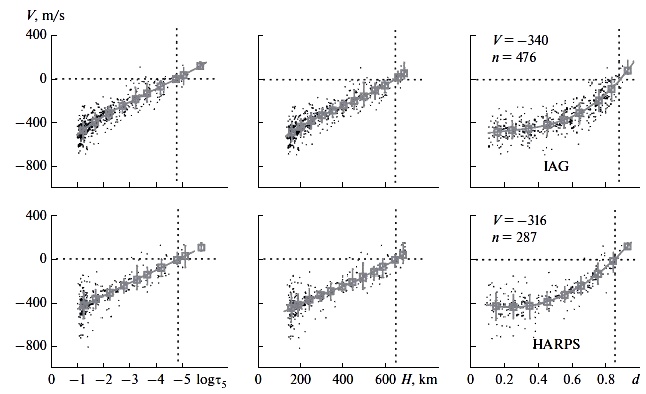}
}
  \caption {\small
Plots of shifts versus optical depth $\log \tau_5$, geometric height $H$, and line depth $d$ obtained for solar lines from the IAG atlas (top panel) and HARPS atlas (bottom panel). The dots are the shifts of individual lines; the gray squares with vertical lines are the mean values and errors in each bin. The gray curves are third-order polynomial approximations. The average granulation shift $V$ (in m/s) and the number $n$ of the lines are shown at the top.
 }
 \end{figure}

%%%%%%%%%%%%%%%%%%%%%%%%%%%%%%%%%%%%%%%%%%%

\section{Results and discussion}

In this study, we focused our attention on the dependence of convective shifts on optical depth in the atmosphere, which directly demonstrates the variation in the granulation velocity through the photosphere. Additionally, we also considered the dependences of the shifts on the line depth since they are often used in the literature to interpret convective shifts and may be of interest for comparison. As for the dependence of the shifts on the equivalent width, excitation potential, and wavelength, their character and properties are similar to the dependences obtained earlier, so we do not present them here

%%%%%%%%%%%%%%%%%%%%%%%%%%%%%%%%%%%%%%%%%%% Figure 3
 \begin{figure}[t]
 \centerline{
 \includegraphics   [scale=1.2]{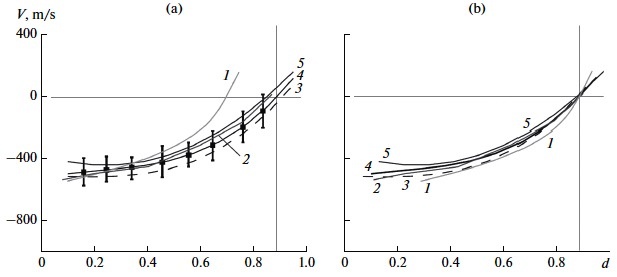}
}
 \caption {\small
Convective shift curves obtained for the Sun in
\cite{2009ApJ...697.1032G, 1999PASP..111.1132H} (curves 1, 2), 
\cite{2016A&A...587A..65R} (curve 3), this analysis, the HARPS atlas (curve 4), and the IAG atlas (curve 5). All curves in panel (a) are represented in their own shift scales; in panel (b), they are shifted so that the zero shift corresponds to the line depth $d=0.88$  (vertical line).
 }
 \end{figure}

%%%%%%%%%%%%%%%%%%%%%%%%%%%%%%%%%%%%%%%%%%%

{\bf Solar convective shift curves.}
Figure 2 shows the solar line shifts obtained for 476 lines from the IAG atlas ($R\approx1000000$) and 287 lines from the HARPS atlas ($R\approx120000$) in scales of optical depths, geometric heights, and line depths. The plots of the shifts versus optical depth and geometric height $H$ in the atmosphere show that the relationship between $\log \tau_5$ and $H$ is almost proportional. The geometric scale of heights is given in kilometers above  $\tau_5 = 1$. With the help of these graphs, we determined the depth of zero shifts to establish the zero point on the scale of shifts in stars. It is equal to  $\log \tau_5=-4.85$ or  $H= 650$~km  and does not depend on the spectral resolution. To determine the zero point of the shift scale, one can also use the dependence of the shifts on the line depth $d$, as was done in  \cite{2009ApJ...697.1032G}. Our results show that, in this case, the zero point will depend on the spectral resolution. We obtained that the zero point corresponds to the line depths $d= 0.88$ for IAG and 0.85 for HARPS. This means that an error of approximately 70~m/s on the shift scale can be allowed due to the lower resolution of the HARPS atlas. Figure 2 shows that the scatter of individual values in the plot of shifts versus line depth $d$ is greater than in the plot versus optical depth $\log \tau_5$. This is due to the influence of the wavelength and excitation potential on $d$. In general, the amount of scatter in each bin is mainly due to observation noise, the presence of invisible blends, laboratory wavelength errors, and different number of lines. We obtained an average convective shift of  $-340\pm159 $ m/s (IAG) and $-316\pm165 $ m/s (HARPS). The slight difference (24 m/s) is due to the different spectral resolution. It is a known fact that the line shifts decrease with the deterioration of the spectral resolution
\cite{2013A&A...550A.103A}. % Allende Prieto, A\&A 550, A103 (2013)).
Our results are in satisfactory agreement with the results of 
\cite{1999ASPC..185..268D, 2017A&A...597A..52M} % Dravins(1999), Meunier et al. A\&A 597, A52 (2017)
($-300$ and $-355$ m/s, respectively).

We obtained the following expressions for the dependence of solar convective shifts $V$ (m/s) on the optical depth $\log \tau_5$ according to the HARPS atlas:
\[V = -601.2 -145.9 \log \tau_5 -8.7 ({\log \tau_5})^2 -0.8 ({\log \tau_5})^3,\]
and according to the IAG atlas:
\[V = -737.8 -245.4  \log \tau_5-32.1 ({\log \tau_5})^2 -2.7 ({\log \tau_5})^3,\]
and for dependences on the line depth, respectively:
\[V=-369.7   -550.3d + 1105.5d^2+70.4 d^3,\]
\[V=-518.6 +  326.7d -783.5d^2+1211.4 d^3.\]\\
These expressions can be used to construct standard convective velocity curves for the Sun as a star. In Fig. 3a, we compared our standard curves with the curves obtained earlier
\cite{2009ApJ...697.1032G, 1999PASP..111.1132H, 2016A&A...587A..65R}.
% (Hamilton  Lester (1999), Gray (2009), Reiners et al.(2016)).
In general terms, they are similar but displaced relative to each other within 100 m/s, except for the curve
\cite{2009ApJ...697.1032G}, % Gray (2009),
which is displaced much more. The differences in the zero point of the shift scale can be caused by errors in the calibration of the wavelength scale, the choice of spectral lines, different methods of measuring the position of the line core, etc. If we combine all the curves as shown in Fig. 3b, it can be seen that they demonstrate satisfactory agreement with each other within the limits of the permissible errors of this analysis.

{\bf Stellar convective shift curves.} Figures 4--6 show the results of measurements of convective shifts for stars (curve 1) in comparison with the standard curve (curve 2) of the shifts obtained for the Sun from the HARPS atlas. First of all, we should note that the number of individual shifts is significantly smaller for the hottest stars HD 189627, HD 102361, and HD 147873 with large  $v \sin i= 5.5$, 5.6, 6.5 km/s, respectively (Fig. 4). This is due to the weakening of the line profiles, which become shallower, broader, and more noisy. As a result, the number of lines suitable for analysis decreases and the variance increases. The second aspect that should be emphasized is the change in the curvature of the obtained dependences and their slope with a change in the effective temperature and rotation velocity. This can be clearly seen in comparison with the solar shift curve. The character of the curves between $\log \tau_5= -1$ and $-3$ ($H=150$ and 350 km) is almost linear but the shift gradient changes at higher heights depending on the temperature of the star. It is larger in stars hotter than the Sun (Figs. 4, 5) and much smaller in cooler stars (Fig.~6). The linear character of the height dependence of the shifts in the lower layers of the photosphere was previously noted for solar lines in the study of granular velocities
\cite{1980ApJ...237.1024K}.%(Keil

In Figs. 4--6, we can see the tendency for convective shifts to change with height, which is common for all stars. In the group of cool K stars (Fig. 6), blue shifts decrease from $-250$ m/s to zero on average in the layers from   $\log \tau_5=-1$ to $-4$. In the group of hotter FG stars (Figs. 4, 5), the temperature of which is 1000 K higher on average, the magnitude of the granulation velocities and the rate of variation are higher. On average, they vary from $-700$ m/s to 0 m/s in the layers from   $\log \tau_5=-1$ to $-5$. Above these layers, we see red shifts in all the stars. They increase in a rather narrow layer of the lower chromosphere from   $\log \tau_5=-5$ to $-5.5$ or $-6$. In cold stars, they reach values of 100 m/s, while they are 300--500 m/s in hot stars. Our results convincingly show that granulation velocities change their direction in the lower layers of the chromosphere.

%%%%%%%%%%%%%%%%%%%%%%%%%%%%%%%%%%%%%%%%%%% Figure 4
 \begin{figure}[t]
 \centerline{
 \includegraphics   [scale=1.1]{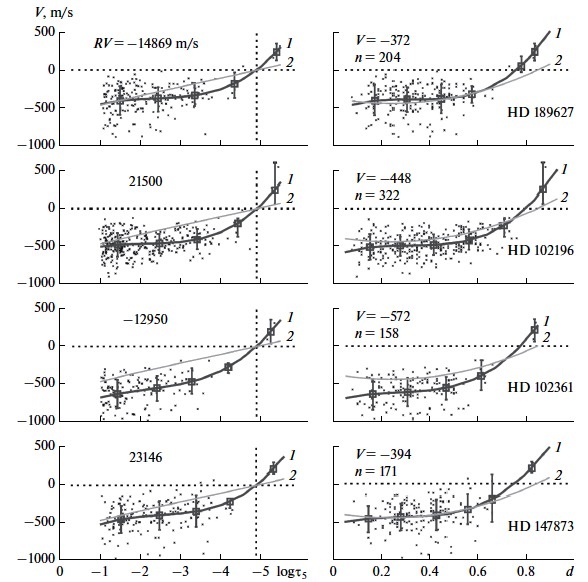}
}
  \caption {\small
Plots of the convective line shifts versus optical depth  $\log \tau_5$ and  line depth $d$ in the spectra of hotter stars (the effective temperature decreases from 6200 to 6000 K top to bottom) with rotation velocities  $v\sin i = 3.5$--6.5  km/s. The dots are the shifts of individual lines, the squares with vertical lines are the mean values and errors in each bin. Curve 1 is the polynomial approximation of the third order; curve 2 shows convective shifts for the Sun. The average convective velocity $V$ over all lines, the number $n$ of the lines, and the derived radial velocity $RV$ for each star are indicated on the panels.
 }
 \end{figure}
%%%%%%%%%%%%%%%%%%%%%%%%%%%%%%%%%%%%%%%%%%%

%%%%%%%%%%%%%%%%%%%%%%%%%%%%%%%%%%%%%%%%%%% Figure 5
 \begin{figure}[!t]
 \centerline{
 \includegraphics   [scale=1.1]{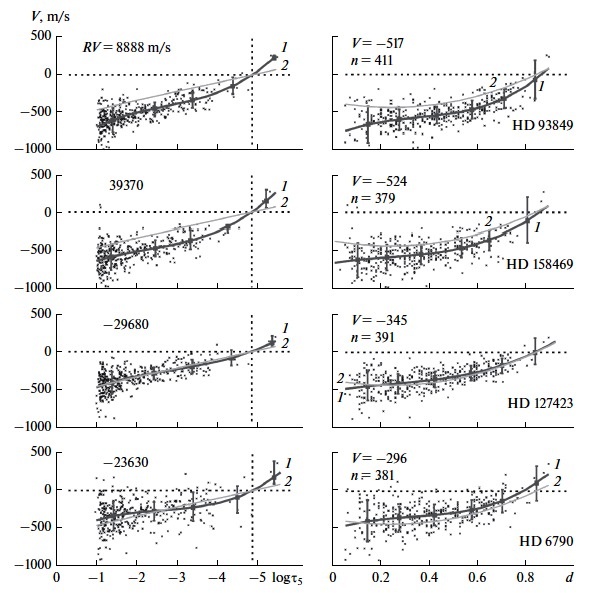}
}
  \caption {\small
Same as Fig. 4 for hotter (6100--6000 K) stars with rotation velocity $v\sin i = 2.5$--3.1~km/s.
 }
 \end{figure}
%%%%%%%%%%%%%%%%%%%%%%%%%%%%%%%%%%%%%%%%%%%

%%%%%%%%%%%%%%%%%%%%%%%%%%%%%%%%%%%%%%%%%%% Figure 6
 \begin{figure}[t]
 \centerline{
 \includegraphics   [scale=1.1]{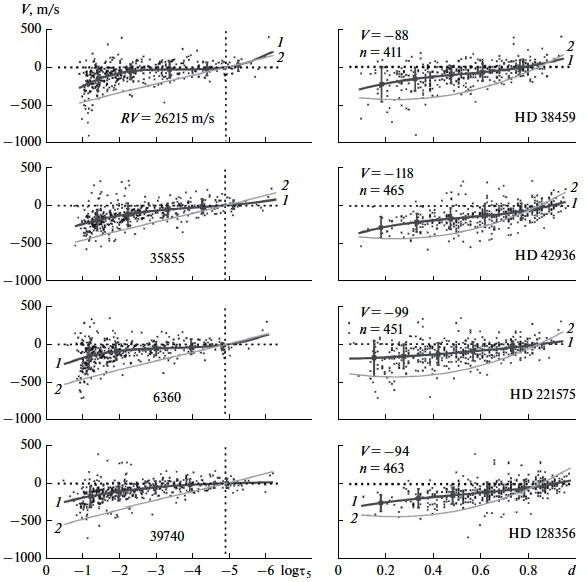}
 }
  \caption {\small
Same as Fig. 4 for colder (5200--4900 K) stars with rotational velocity $v\sin i = 1.0$--1.9~km/c.
 }
 \end{figure}
%%%%%%%%%%%%%%%%%%%%%%%%%%%%%%%%%%%%%%%%%%%

%%%%%%%%%%%%%%%%%%%%%%%%%%%%%%%%
%%==========================================================

%___________________________________ Table 2
\begin{table}
\centering
 \caption{\small Results of measuring the convective velocity $V$ averaged over $n$ lines, as well as the radial velocity  $RV$,  obtained by us, the radial velocity from the SIMBAD base $RV_{SIM}$, the data on the micro $\xi$  and macroturbulent $\zeta$ velocities and rotation velocities  $v \sin i$.}
 \vspace {0.3 cm}
\label{T:2}

\footnotesize
\begin{tabular}{lcccccccccc}
\hline\hline
HD    &Type & $T_{\rm eff}$& $V$ & $rms$   &$n$ &$RV$   &$RV_{SIM}$&$\xi$ &$\zeta$ &$v \sin i$ \\
      &     &   K          &(m/s)&       &    & (km/s)& (km/s)   &(km/s)& (km/s) & (km/s)    \\
\hline
6790  & G0 V      &6012   &$-$296    &190  &381 &-23.630&   $-$23.630 &0.8   &3.2     &2.9        \\
38459 & K1 IV-V   &5233   &$-$88     &149  &436 & 26.215&    26.552 &1.0   &3.2     &1.9       \\
42936 & K0 IV-V   &5126   &$-$118    &144  &469 & 35.855&    34.080 &0.7   &1.7     &1.0       \\
93849 & G0/1 V    &6153   &$-$517    &192  &411 &  8.888&     8.548 &1.2   &2.9     &3.1       \\
102196& G2 V      &6012   &$-$448    &163  &322 & 21.500&    21.460 &1.4   &4.3     &3.6       \\
102361& F8 V      &5978   &$-$572    &209  &158 &-12.950&     8.100 &1.4   &5.6     &5.0       \\
127423& G0 V      &6020   &$-$345    &156  &391 &-29.680&   $-$29.700 &1.0   &2.9     &2.5       \\
128356& K2.5 IV   &4875   &$-$94     &126  &463 & 39.740&    39.918 &0.7   &1.7     &1.0       \\
147873& G1 V      &5972   &$-$394    &202  &171 & 23.146&    22.916 &1.5   &6.0     &6.5       \\
158469& F8/G2 V   &6105   &$-$524    &184  &379 & 39.370&    39.250 &1.2   &3.6     &3.1       \\
189627& F7 V      &6210   &$-$372    &187  &204 &-14.869&   $-$14.989 &1.5   &5.5     &5.9       \\
221575& K2 V      &5037   &$-$99     &147  &451 &  6.360&     6.600 &0.9   &2.8     &1.9       \\
Sun~~~& G2 V      &5777   &$-$316    &165  &287 &   --  &      --   &0.8   &2.1     &1.8       \\
Sun(IAG)&         &5770   &$-$340    &159  &476 &   --  &      --   &0.8   &2.1     &1.8       \\
\hline
\end{tabular}
\end{table}
\noindent

%%%%%%%%%%%%%%%%%%%%%%%%%%%%%%%%%%%%%%%%%%% Figure 7
 \begin{figure}[t]
 \centerline{
 \includegraphics   [scale=0.9]{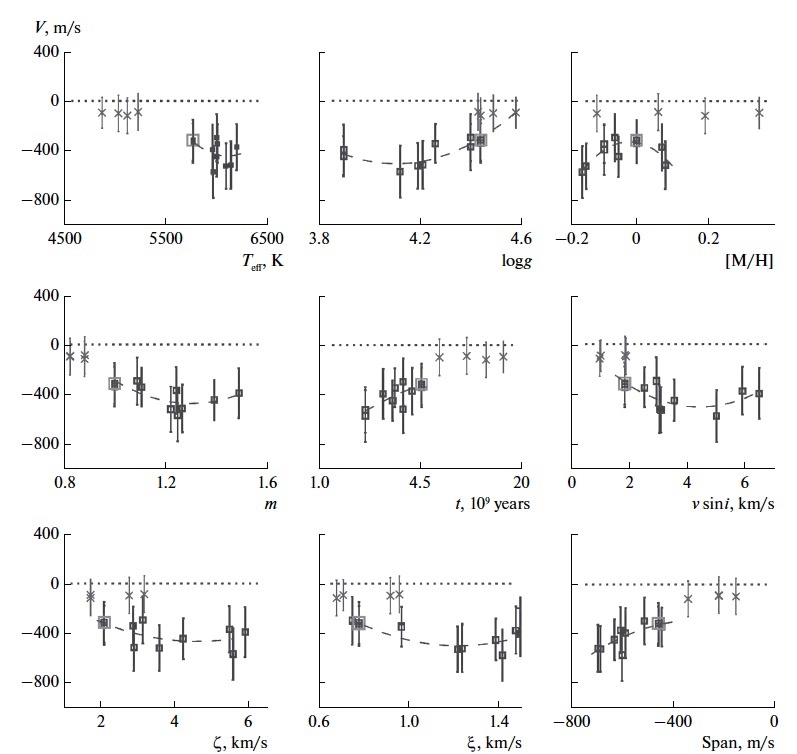}
}
  \caption {\small
Average convective velocity of stars obtained from the average value of the shifts of all lines depending on the effective temperature, surface gravity, metallicity, mass, age, the line-of-sight projection of the rotation velocity, macro- and microturbulent velocities, and the span of the mean bisector of the lines (the black squares are for the group of hotter stars, oblique crosses are for cooler stars, the large gray square is for the Sun, the dashed line is an approximation curve for the hotter group of stars).
 }
 \end{figure}
%%%%%%%%%%%%%%%%%%%%%%%%%%%%%%%%%%%%%%%%%%%

%%%%%%%%%%%%%%%%%%%%%%%%%%%%%%%%%%%%%%%%%%% Figure 8
 \begin{figure}[t]
 \centerline{
 \includegraphics   [scale=1.1]{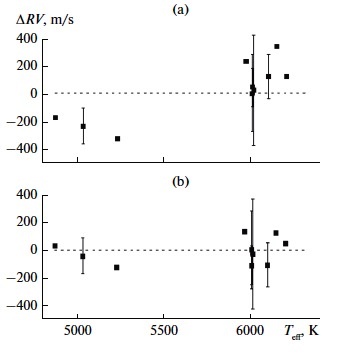}
}
  \caption {\small
Difference of the radial velocities of stars between the $RV$ values obtained in this analysis and  $RV_{SIM}$  values depending on the effective temperature: (a) for the original data from the SIMBAD database and (b) for the data from the SIMBAD database corrected for convective velocities derived in this analysis.
 }
 \end{figure}
%%%%%%%%%%%%%%%%%%%%%%%%%%%%%%%%%%%%%%%%%%%

For stars, the average convective velocity obtained from the average value of the shifts of all lines is presented in Table 2 and in Fig. 7 depending on the characteristic parameters of the star. The dependence of the convective velocity on the effective temperature is most pronounced in comparison with others. It is remarkably confirmed by the results
\cite{2017A&A...597A..52M} % Meunier et al.A\&A 597, A52 (2017)
for GK dwarfs. There is also a tendency to an increase in the convective velocity in stars with lower metallicity and gravity, greater mass, lower age, and higher rotation velocity. The first sign of granulation (macro- and microturbulent velocities) and the second sign (asymmetry of the lines) noticeably correlate with convective velocity. Here, the asymmetry of the lines is represented by a quantitative characteristic called the span of the average bisector. We measured the macro- and microturbulent velocities and the bisector span for these stars in our previous study
\cite{2020KPCB...36..291S}. % (Шеминовой (2020).

The behavior (Fig. 7) of the dependences on the star's parameters is explained by the physical processes of decay of convective motions in the photosphere, which were previously discussed in
\cite{2002ApJ...566L..93A, 2009ApJ...697.1032G, 2017A&A...607A.124M, 2008A&A...492..841R}.
% Gray,  Allende Prieto, Ramirez, Meunier и др.
Undoubtedly, the effective temperature plays a decisive role in the development of stellar granulation. The main cause of granulation is the flow of energy coming from the star's interior. The higher the temperature, the more energy is spent on transporting the radiation flux to the surface of the star, the stronger the granulation. The increase in the slope of the convective shift curves (Figs.~4--6) with increasing effective temperature is partly due to a general drop in opacity due to a decrease in absorption by the negative hydrogen ion, and, therefore, deeper layers with higher velocities can be observed. Lower gravity and lower metallicity result in a more extended and more transparent photosphere, which also allows higher velocities to be seen. The higher the effective temperature and the lower the gravity and metallicity, the greater the force of penetrating convection will be, and the average convective shift of the lines will also be larger. The temperature gradient also affects granulation
\cite{2013A&A...558A..49B, 1998ApJ...499..914S}.
%(stein, nordlund (the astrophysical journal, 499,914-933, 1998), Beeck A\&A 558, A49 (2013)).
The larger it is, the higher temperature is achieved at a shorter geometric distance, so the optical depth will be smaller, and, therefore, we observe deeper layers with a higher gas temperature and higher convective velocities. The temperature of the star also affects the structuring of the stellar atmosphere
\cite{2013A&A...558A..49B}. % Beeck AA 558, A49 (2013).
The brighter the granule, the higher its velocity. Large granules have a lower velocity on average than small granules. The average granule size decreases for cooler, more compact stars, i.e., the granules of the main sequence stars become smaller as the effective temperature decreases and the gravitational acceleration increases. 

A deeper understanding of the character of the height dependences of convective velocities is given by the results of studying granulation in quiet regions of the Sun with a high spatial resolution (approximately  $0.5^{\prime\prime}$). It was shown in a number of studies
\cite{2015KPCB...31...65B, 1995A&AS..109...79E, 2005A&A...441.1157P}
 that temperature inversion in granules and intergranules occurs at heights of 200--300~km and higher. These are the heights at which the height dependences of the shifts we obtained deviate from the linear dependence. According to the data in
\cite{2015KPCB...31...65B},  %Баран, Стодилка (2015)
the pattern of the processes of granulation is as follows. The substance of the central parts of the granules becomes colder at these heights and continues to move upward. This is due to the rapid adiabatic expansion and radiative cooling of the ascending flow, heating due to the compression of the descending gas, and the sensitivity of the absorption coefficient of the negative hydrogen ion to temperature variations. The height of the inversion strongly depends on the temperature contrast and velocity in granules and intergranules at the continuum level. The larger they are, the higher the height at which the inversion takes place. In addition to the temperature inversion, velocity inversion occurs. Approximately 12\% of granules and intergranules change the direction of movement to the opposite in the middle photosphere, and this occurs in most of the granulation pattern above 500~km. Despite the inversion of temperature and velocity, more than 40\% of the granulation structure is retained in the form of columns up to an height of 650 km and higher
\cite{2009A&A...506.1405K, 2011A&A...531A..17R}.
%(Kostik R., Khomenko E., Shchukina N. AA.2009.506, N 3.P. 1405-1414), Rutten .
It should be noted that these are the heights at which the core of the very strong line Mg I 517.2 nm forms.  Only large granules with a diameter of more than 1500 km reach heights of 650--700 km. Although the granulation structure still exists here, it differs significantly from the granulation pattern seen in the wings of the Mg I 517.2 nm line
\cite{2011A&A...531A..17R}. %(Rutten).
These observational facts are confirmed by the red shifts of very strong lines we obtained at heights above 600 km.

It should be emphasized that the red shifts were predicted in studies on the numerical three-dimensional modeling of convection in the near-surface layers of stars (e.g.,
\cite{2013A&A...550A.103A, 2009A&A...501.1087R}).
%Ramirez A\&A 501, 1087-1101 (2009)), Allende Prieto A\&A 550, A103 (2013)).
According to the simulation results, granulation on the stellar surface occurs in the region of penetrating convection, i.e., in the region of ejections of hot matter from the subphotospheric layers. Due to the forces of buoyancy, the substance that rises in the granules moves upward by inertia and slows down with height until it spends its energy. Then the direction of this motion changes to the opposite and the substance begins to descend down the colder intergranular gaps. The region where the rearrangement of the granulation pattern takes place is called the region of reverse granulation. The change in the direction of granulation velocities causes red shifts of very strong lines, the cores of which form in the region of reverse granulation at heights above 600 km. According to modeling
\cite{2009A&A...501.1087R}, %Ramirez A\&A 501, 1087-1101 (2009))
red shifts reach 100 m/s for Fe I lines in the atmosphere of a K dwarf, while the maximum blue shifts are approximately $-200$ m/s. According to our results, the strongest Mg I lines show the largest red shifts in all solar-type stars. The higher the effective temperature, the greater the red shift. For cool stars, the red shift of the strongest Mg I line is 100 m/s, that for the Sun is 160 m/s, and that for hot stars reaches 300 m/s. Hence, it follows that strong Mg I lines cannot be used as reference points for constructing the absolute scale of shifts in stars.

{\bf Radial velocities of the stars.} As a byproduct of this analysis, the radial velocities $RV$ of the stars were obtained. They are presented in Table 2, and the differences between them and the SIMBAD database  $RV_{SIM}$ are shown in Fig. 8a depending on the effective temperature. Large deviations of the radial velocities were found for the stars HD 102361 and HD 42936: $-21050$ and 1775 m/s, respectively. The cause of these deviations is not known. For the remaining ten stars, the difference does not exceed $\pm340$ m/s. As seen from Fig. 8a, the obtained differences are negative for cool stars and positive for hot stars. This suggests that these deviations are possibly caused by the unaccounted temperature dependence of the convective velocities of the stars in the SIMBAD data, i.e., the same convective velocity obtained for the Sun ($-300$ m/s) was taken into account for all solar-type stars. To verify this, we introduced our convective velocities $V$ in $RV_{SIM}$:
 \[RV_{SIM}^{cor} = RV_{SIM} -300-V.\]
The differences between our $RV$ and the corrected radial velocities SIMBAD $RV_{SIM}^{cor}$ are shown in Fig. 8b. As can be seen, the obtained differences do not now depend on temperature and do not exceed $\pm 125$  m/s, and their average absolute difference is 75 m/s, which does not exceed the error limits for this analysis. This implies the following: (1) the dependence of the convective velocities of stars on the effective temperature cannot be neglected when determining the radial velocities by spectroscopic methods and (2) the radial velocities of stars we derived are more accurate than the SIMBAD data. Their error does not exceed the root mean square errors of convective velocities presented in Table 2.

\section{Conclusions}

For a small sample of FGK dwarfs with effective temperatures in the range of 6200--4900~K, we measured the iron line shifts in the observed spectra with a resolution $R \approx 120 000$, signal-to-noise ratio S/N~$ > 100$, and spectral coverage from 410 to 680~nm. The number of lines used for the analysis was from 158 to 476, depending on the star. To determine convective shifts, absolute scales were constructed based on the graphs of convective line shifts in the solar flux spectrum versus optical depth, assuming that the zero shift of lines in the spectra of the Sun and solar-type stars occurs at the same optical depth. Since zero shifts of iron lines are not observed for stars hotter than the Sun with a rotation velocity of more than 5 km/s, we have added to the lists of lines several very strong Ca I and Mg~I lines, which form in the lower chromosphere and show zero and red shifts. This made it possible to expand the range of shifts of the observed lines and determine the zero point of the shift scale more reliably.

Our results confirmed the typical granulation properties in solar-type stars. The slope of the dependences of convective shifts increases for hotter stars. Weak lines that form in deep regions of the photosphere show blue shifts from $-700$ m/s in the hottest stars to $-250$ m/s in the coolest stars of our sample. Strong Fe I and Ca I lines show zero shifts in the spectra of the Sun and in cooler stars, and very strong Mg I lines show red shifts in all stars. Red shifts increase with the temperature of the star and reach maximum values of approximately 340 m/s in the hottest stars. The convective shift averaged over all the lines increases from $-100$ to $-560$ m/s with an increase in effective temperature from 5000 to 6000 K and a decrease in metallicity, surface gravity, and age of the star. The average convective shift correlates with the rotation velocity of the star as well as with micro- and macroturbulent velocities and asymmetry of the lines. All these properties are consistent with the fact that the strength of granulation in stars is mainly determined by temperature. The higher it is, the higher the granulation velocities and the greater their gradient.

The derived radial velocities, with the exception of two stars HD 102361 and HD 42936, are in satisfactory agreement with the SIMBAD data if the latter are corrected for the value of convective velocities that depend on the effective temperature of the star. The cause of the large difference for HD 102361 (-21050 m/s) and HD 42936 (1775 m/s) is not known. This question is still open and requires further study. The error in our radial velocities does not exceed the mean root-mean-square error in bins for the derived convective velocities in stars, which is $\pm$141 m/s for cooler stars and $\pm181 $ m/s for hotter stars.

Thus our results indicate that the average convective velocities in solar-type stars can be quite significant from  $-300$ to $-560$~m/s s in FG stars with an effective temperature of 6000--6200 K and from  $-90$ to $-120$~m/s in K stars with a temperature of 4900--5200 K. In this regard, measurements of the absolute radial velocities should take into account the individual convective velocities of the star. For more accurate measurements of convective velocities and a deeper understanding of the relationship of granulation with other stellar phenomena, as well as an understanding of the influence of surface convection on the structure and evolution of stars, it is necessary to measure and trace the line shifts for a larger number of stellar spectra with higher quality. 

\vspace {1cm}  

{\bf Acknowledgments.}
I am sincerely grateful to Ya. Pavlenko and A. Ivanyuk for providing the observed spectra of the stars and discussing
the results.

\vspace {1cm}
{\bf Conflict of interests.}
The author declares that she has no conflict of interests.

\end{document}